# RANDOMIZING THE SUPERSTRING

W. Siegel[1]

*Institute for Theoretical Physics*
*State University of New York, Stony Brook, NY 11794-3840*

**ABSTRACT**

We find a simpler formulation of the Green-Schwarz action, for which the Wess-Zumino term is the square of supersymmetric currents, like the rest of the action. On a random lattice it gives Feynman diagrams of a particle superfield theory.

---

[1] Internet address: siegel@insti.physics.sunysb.edu.

# 1. INTRODUCTION

Green and Schwarz [1] found the first action for the mechanics of the superstring that was invariant under spacetime supersymmetry and $\kappa$ symmetry [2] (needed for gauging away half of the anticommuting coordinates). In an effort to replace the troublesome second-class constraints (needed for eliminating half of the remaining fermionic coordinates), we introduced the affine Lie algebra of supersymmetry covariant derivatives $D_\alpha, P_a, \Omega^\alpha$ [3] and wrote a complete set of first-class constraints (and corresponding action) in terms of them [4]. Like the original Green-Schwarz formulation, the only coordinates appearing (ignoring Lagrange multiplier gauge fields like the world-sheet metric) were those familiar from the usual superspace applied to particles, $\theta^\alpha$ and $x^a$ (now defined on the world sheet rather than the world line). Unfortunately, elimination of second-class constraints did not solve the covariant quantization problem [5], and it seemed that additional coordinates were necessary [6].

Green, to try to fix the first-class formalism, derived a new representation of the supersymmetric affine Lie algebra with an additional spinor coordinate [7] by applying the usual group theoretic construction [8], which uses separate coordinates $\theta^\alpha, x^a, \phi_\alpha$ for all the generators of the algebra. This construction actually produces two such algebras, one for each chirality. The only available expressions for the supersymmetry generators were the zero modes of the generators of the affine Lie algebra. Unfortunately, one chirality of these has the wrong sign in its algebra for unitarity. This follows from the usual argument that the square of supersymmetry must be plus the energy. This was to be expected from the fact that the generators of the affine Lie algebra are to be identified with the supersymmetry covariant derivatives, which have this "wrong" sign even for the superparticle. We will show below how the correct interpretation of the second chiral algebra reproduces the original representation, and gives a more direct derivation of it.

Another approach to string quantization, that directly expresses the string as a bound state of particles, is the random lattice [9]. There the world-sheet metric is replaced by a sum over lattices with the geometry of Feynman graphs, the surface of the string resulting from the usual limit of the 1/N expansion. This method was applied to the Green-Schwarz superstring [10], but the Wess-Zumino term led to problems when relating to the corresponding superparticle theory: (1) When defined on the Feynman diagram lattice, this term was defined only on loops, which has no analog in Feynman rules; and (2) the Green-Schwarz action is supersymmetric only up to a surface term, which is sufficient to make each Feynman diagram supersymmetric, but not the individual vertices. We will show that the use of the redundant coordinate $\phi_\alpha$ yields a simple reformulation of the second-class action in terms of just (supersymmetry) covariant momenta, avoiding the problems the Green-Schwarz action had on the random lattice. This method of quantization might also shed some light on the quantization of superparticles, just as string field theory has given new understandings of particle field theory.

In the following section we will review the construction of representations of Lie (super)algebras on group space, the corresponding actions for particle mechanics, and representations of affine Lie (super)algebras and their string mechanics actions. In section 3 we will discuss the quantization of such strings on random lattices. We specialize these methods to the superstring in section 4. The main results of this paper are described in the final section. They include very simple forms for the action of the superstring: The second-order Lagrangian is just

$$L = -\tfrac{1}{2}\mathbf{g}^{mn} J_n^a J_{ma} - i\tfrac{1}{2}\epsilon^{mn} J_n^\alpha J_{m\alpha}$$

in terms of the currents

$$J^\alpha = \partial\theta^\alpha, \quad J^a = \partial x^a - i(\partial\theta^\alpha)\gamma^a_{\alpha\beta}\theta^\beta, \quad J_\alpha = \partial\phi_\alpha - 2i(\partial x^a)\gamma_{a\alpha\beta}\theta^\beta - \tfrac{2}{3}(\partial\theta^\beta)\gamma^a_{\beta\gamma}\theta^\gamma\gamma_{a\alpha\delta}\theta^\delta$$

invariant under the supersymmetry transformation

$$\delta\theta^\alpha = \epsilon^\alpha, \quad \delta x^a = -i\epsilon^\alpha\gamma^a_{\alpha\beta}\theta^\beta, \quad \delta\phi_\alpha = 2i\epsilon^\beta\gamma_{a\alpha\beta}x^a + \tfrac{2}{3}(\epsilon^\beta\gamma^a_{\beta\gamma}\theta^\gamma)\gamma_{a\alpha\delta}\theta^\delta$$

($\mathbf{g}^{mn} \equiv \sqrt{-g}g^{mn}, \epsilon_{01} = 1$.) Up to a total derivative, this Lagrangian is identical to the Green-Schwarz Lagrangian. However, that derivative term contains all the $\phi$ dependence. The existence of the coordinate $\phi$ allows the Wess-Zumino term to be expressed in the manifestly supersymmetric way above (without the extension of the world sheet to a three-dimensional space). It also allows the Lagrangian to be expressed completely in first-order form as

$$L = iJ_m^A P_A^m - \tfrac{1}{2}\mathbf{g}_{mn} P^{na}P_a^m + 2i\epsilon_{mn} P^{n\alpha}P_\alpha^m$$

where $P_A^m$ are the covariant (group-space) derivatives $D, P, \Omega$, represented as independent variables. Consequently, the action can be put on a lattice in a way that corresponds to Feynman rules derivable from the field theory action for a superparticle. The fact that the Wess-Zumino term has no explicit dependence on the coordinates $\theta, x, \phi$, but only on their covariant momenta $D, P, \Omega$, means that the kinetic and interaction terms of this field theory are independently supersymmetric; i.e., the supersymmetry transformations are homogeneous in the fields (as in usual superfield theories).

## 2. REVIEW OF SUPERALGEBRAS

A (super)group element $U$ can be parametrized on the group space, e.g., by the usual exponential parametrization

$$U = e^{-iX^A T_A}$$

in terms of group-space coordinates $X^A$ and Lie algebra generators $T_A$, which satisfy

$$[T_A, T_B\} = if_{AB}{}^C T_C$$

We can also define a group metric (not necessarily the Cartan metric, which may be degenerate):

$$tr(T_A T_B) = \eta_{AB}$$

The action of the group on these coordinates can be described in two different ways, corresponding to left and right group multiplication:

$$U' = U_0 U, \quad \tilde{U}' = U U_0$$

where $U_0$ is the transformation, parametrized in a similar way (so $X_0^A$ are the parameters of the transformation). Because left and right multiplication commute (associativity), so do the two sets of transformations. We can then define combinations of two different coordinate points that are invariant under one or the other of these two transformations:

$$U_{12} \equiv U_2^{-1} U_1 \quad \Rightarrow \quad U_{12}' = U_{12}, \quad U_{12} = U_{21}^{-1} = e^{-iX_{12}^A T_A}$$

(and similarly for the other handedness). We then have $X_{12} = -X_{21} = X_1 - X_2 + ...$, where the "..." covariantizes this finite difference. These finite differences appear in propagators, and as the lattice versions of the corresponding infinitesimal differences, as we will see in the next section.

The infinitesimal forms of these differences are

$$U^{-1} dU = -i(dX^M) E_M{}^A T_A, \quad (dU)U^{-1} = -i(dX^M) \tilde{E}_M{}^A T_A$$

which define the covariant derivatives

$$P_A = E_A{}^M i\partial_M, \quad \tilde{P}_A = \tilde{E}_A{}^M i\partial_M$$

The covariant derivatives $P$ and $-\tilde{P}$ are two group-space coordinate representations of the generators $T_A$:

$$[P_A, P_B\} = if_{AB}{}^C P_C, \quad [\tilde{P}_A, \tilde{P}_B\} = -if_{AB}{}^C \tilde{P}_C, \quad [P, \tilde{P}\} = 0$$

Expanding the identities
$$U^{-1}(P_A U) = (\tilde{P}_A U)U^{-1} = T_A$$
in $T$, we also find
$$P_{1A} X_{12}^B = -P_{2A} X_{12}^B = i\delta_A^B$$

These group elements have a quantum mechanical description: Defining the currents
$$J = iU^{-1}\dot{U}, \quad \tilde{J} = i\dot{U}U^{-1}$$
we can consider the second-order Lagrangian
$$L_2 = -\tfrac{1}{2} tr\, J^2 = -\tfrac{1}{2} tr\, \tilde{J}^2 = -\tfrac{1}{2} \dot{X}^M \dot{X}^N G_{NM}$$
where
$$G_{MN} = E_M{}^A E_N{}^B \eta_{BA} = \tilde{E}_M{}^A \tilde{E}_N{}^B \eta_{BA}$$
is the metric on the group space. (We are now a bit sloppy with signs corresponding to supergroups, but they are determined by the ordering of contracted indices.) If $\eta$ is invertible, we can write the first-order form
$$L_1 = tr(iJP - \tfrac{1}{2}P^2) = tr(i\tilde{J}\tilde{P} - \tfrac{1}{2}\tilde{P}^2)$$
where $P = P^A T_A$ (and similarly for $\tilde{P}$), and $P_A$ takes the above form of covariant derivatives upon quantization. (The $i$'s are for Euclidean time, functionally integrating $e^{\int dt\, L}$.)

More generally, we can consider first-order Lagrangians of the form
$$L_1 = tr(iJP) - H(P)$$
where the Hamiltonian $H$ corresponds to any Casimir of the group: i.e., $[H, P_A] = 0$. (Note that $H(P)$ has no $T$ dependence; this commutator invloves $[P_A, P_B\}$, not $[T_A, T_B\}$.) Then $P$ satisfies the equation of motion $\dot{P} = 0$. As a result, we can perform the first-quantized path integral exactly: With appropriate measure factors,
$$\int \mathcal{D}U\, \mathcal{D}P\, e^{\int dt\, L_1} = e^{-(t_f - t_i)H(P_i)} \int dP\, (U_f^{-1} U_i)|_{T=P}$$
$$= \int dP\, e^{-iX_{ij}^A P_A - (t_f - t_i)H(P)}$$
where $P$ is now a single remaining integration variable (since $\dot{P} = 0$). This result is equivalent to writing the propagator in operator notation
$$|X\rangle = U(X)|0\rangle \quad \Rightarrow \quad \Delta = \langle 0|U_f^{-1} e^{-(t_f - t_i)H(T)} U_i|0\rangle$$

$$H(\tilde{P}) = H(P) \quad \Rightarrow \quad \Delta = e^{-(t_f - t_i)H(P_i)} \langle 0 | U_f^{-1} U_i | 0 \rangle$$

where $|0\rangle$ is the state in the Hilbert space corresponding to the identity element in the group space. In the case of SU(2), these methods correspond to the use of Euler angles to describe the quantization of a rigid body.

In the generalization to affine Lie algebras,

$$J_m = iU^{-1}\partial_m U = (\partial_m X^M) E_M{}^A T_A$$

etc. The second-order action then takes the form

$$S_2 = -tr \left( \int d^2\sigma \tfrac{1}{2}\mathbf{g}^{nm} J_m J_n + i \int d^3\sigma \tfrac{1}{3}\epsilon^{pnm} J_m J_n J_p \right)$$
$$= - \int d^2\sigma \tfrac{1}{2}(\partial_m X^M)(\partial_n X^N)(\mathbf{g}^{nm} G_{NM} + i\epsilon^{nm} B_{NM})$$

where $B$ is the potential for the torsion (structure constants):

$$\tfrac{1}{2}\partial_{[M} B_{NP)} = E_M{}^A E_N{}^B E_P{}^C f_{ABC} = \tilde{E}_M{}^A \tilde{E}_N{}^B \tilde{E}_P{}^C f_{ABC}$$

$P$ and $\tilde{P}$ are now functions of $\sigma$:

$$P_A = E_A{}^M (i\delta_M + X'^N B_{NM}) + X'^M E_M{}^B \eta_{BA}$$
$$\tilde{P}_A = \tilde{E}_A{}^M (i\delta_M + X'^N B_{NM}) - X'^M \tilde{E}_M{}^B \eta_{BA}$$
$$[P_A(1), P_B(2)\} = 2i\delta'(2-1)\eta_{AB} + i\delta(2-1) f_{AB}{}^C P_C$$
$$[\tilde{P}_A(1), \tilde{P}_B(2)\} = -2i\delta'(2-1)\eta_{AB} - i\delta(2-1) f_{AB}{}^C \tilde{P}_C$$

where $\delta_M \equiv \delta/\delta X^M$. (These expressions for $P$ and $\tilde{P}$ are for both the Euclidean and Minkowski world sheets.)

## 3. SUPERRANDOMIZATION

We now consider general properties of the random lattice approach to relating superstrings to superparticles. The main point of the random lattice approach is that the first-quantized path integral for a string on a random lattice is identical to the second-quantized Feynman diagram integral for some particle (preon/parton). This choice of random lattices makes checking unitarity (determining measure factors) and treatment of zero modes (overall momentum conservation $\delta$ functions for each disconnected graph) straightforward, since these properties of the string theory then follow from those of the underlying particle theory. The continuum limit of the lattice string is then identified with the 1/N ("planar") limit of the particle theory, whose fields are NxN matrices.

For the bosonic string, we start with the second-order continuum action

$$S_2 = \int \frac{d^2\sigma}{2\pi} \sqrt{-g}[-\tfrac{1}{2}g^{mn}(\partial_m X)(\partial_n X) + \lambda + (ln\ \zeta)R]$$

where $\lambda$ is the cosmological (Liouville) term and $R$ is the curvature (Euler) term, for constants $\lambda$ and $\zeta$ (in units $\alpha' = 1$). This is discretized as

$$S_2' = -\tfrac{1}{2}\sum_{\langle ij \rangle}(x_i - x_j)^2 + \lambda \sum_i 1 + (ln\ \zeta)Euler$$

where the sums are over propagators (links) $\langle ij \rangle$ and vertices $i$, and "Euler" is the Euler number for that graph (which can be defined as a surface with the help of the 1/N expansion [11]). Then the first-quantized path integral for the string becomes a second-quantized Feynman diagram for a particle, described by an NxN matrix field:

$$\int \mathcal{D}g \int \mathcal{D}x\ e^{S_2} \quad \to \quad \sum_{graphs} \int \prod_i dx_i\ e^{S_2'}$$

where we recognize

$$\Delta(x_i, x_j) = e^{-(x_i-x_j)^2/2},\ G = e^\lambda,\ \zeta = N$$

for the propagator $\Delta$ and coupling (vertex) $G$ for the particle (with the value of $\zeta$ being the usual topological relation for the 1/N expansion). These Feynman rules follow from the particle action (using $e^{S_P}$ in the second-quantized path integral)

$$S_P = tr \int dx \left(-\tfrac{1}{2}\psi e^{-\Box/2}\psi + \frac{1}{n!}\frac{G}{N^{n/2-1}}\psi^n\right)$$

where we have chosen the vertex to be n-point. For purposes of deriving strings in the $N \to \infty$ limit, we assume that we can approximate the kinetic operator as $-e^{-\Box/2} \approx -1 + \tfrac{1}{2}\Box$, the usual one for a massive particle. (However, notice that $G$ is positive, which corresponds to wrong-sign $\psi^4$ theory for $n = 4$, which is asymptotically free in four dimensions, and with the 1/N expansion is "asymptotically convergent" [12].)

The bosonic string is described by an Abelian affine Lie algebra. To generalize to superstrings, which are described by nonabelian ones, we first consider general properties of Feynman graphs for particle field theories described by the corresponding ordinary Lie algebras. The basic idea is that the particle is described by superfields $\psi(X^M)$, and invariant actions are written as integrals $dX$ over this superspace (with group invariant measure) of local Lagrangians expressed in terms of just these fields and their covariant derivatives $P_A$. Variation of such actions then gives propagators that are expressed as covariant derivatives acting on $\delta$ functions:

$$\Delta(X_i, X_j) = f(P_i)\delta(X_i - X_j) = \int dP_{ij}\, f(P_{ij})e^{-iX_{ij}P_{ij}}$$

where $P_{ij}$ are integration variables ($P_i$ are operators) and $X_{ij}$ are the covariant differences of the previous section. A full Feynman supergraph is then given by the product of propagators with covariant vertex operators acting on them:

$$\int \left(\prod_{\langle ij \rangle} dP_{ij}\right)\left(\prod_i{}' dX_i\right) F(P_{ij})e^{-i\sum_{\langle ij \rangle} X_{ij}P_{ij}}$$

where the $P$ integration is over all propagators, and the $X$ integration is over all vertices (but not over $X$'s on the free ends of external lines). For the usual superfields we have coordinates $X^M = (\theta^\alpha, x^a)$, the covariant derivatives $d_\alpha = i\partial/\partial\theta^\alpha - \gamma^a_{\alpha\beta}\theta^\beta\partial/\partial x^a$ and $p_a = i\partial/\partial x^a$, and $X_{ij} = (\theta_i - \theta_j, x_i - x_j - i\theta_i\gamma\theta_j)$. To describe superstrings we will need an additional spinor coordinate $\phi_\alpha$, and perhaps additional coordinates associated with (super)spin and internal symmetry, which may be necessary to cancel conformal anomalies, depending on the spacetime dimension, and whether we want a Liouville term in the continuum limit.

We can identify this as the momentum-space (first-order) form of a first-quantized string path integral if we identify

$$F(P) = e^{-H(P)}$$

$P$ will be identified with string currents of conformal weight one, so $H$ must be quadratic. (In the bosonic case, $F \sim e^{-p^2/2}$.) Because of the results of the previous section, the intermediate step of writing the discretized string is unnecessary: The result of discretizing the $JP$ and $\mathbf{g}PP$ terms in the string mechanics action is to produce a propagator resulting from analogous terms in a first-quantized path integral for a particle:

$$\int d^2\sigma(iJ_m P^m - \tfrac{1}{2}\mathbf{g}_{mn}P^m P^n) \quad \to \quad \int d\tau(iJP - \tfrac{1}{2}P^2)$$

but with the condition $\tau_f - \tau_i = 1$ on the difference between the final and initial times. (The propagator for $\tau_f - \tau_i = 1$ is $e^{-H}$.) This means that the form of the propagator (and similarly for the vertices) can be read off directly from the continuum first-order string mechanics action, essentially by replacing covariant string momenta $P_A^m$ with the corresponding covariant particle derivatives $P_A$, and exponentiating.

We therefore look for an $S_1$ of the generic first-order form

$$S_1 = i(\partial_m X^M) E_M{}^A P_A^m - \tfrac{1}{2}(\mathbf{g}_{mn} G^{AB} - i\epsilon_{mn} B^{AB}) P_B^n P_A^m + ...$$

where "..." is the part independent of the coordinates $X^M$ (cosmological and curvature terms). On the random lattice, $i(\partial X)EP$ becomes $-iX_{ij}P_{ij}$, while the other terms become $-H(P_{ij})$. In general, $G$ and $B$ could be coordinate dependent (but without world-sheet derivatives $\partial_m$, and independent of the momenta $P$). However, group invariance of $S_P$ requires that they be coordinate-independent, since $P$ is by definition invariant. (But coordinate dependence is allowed when background fields are coupled.) While the $(\partial X)EP$ and $\mathbf{g}$ terms determine the propagator as described above, the $\epsilon$ term contributes only to the vertex operators: $\mathbf{g}_{mn}$ acts as the conformal gauge metric (on the *Euclidean* world sheet) $\delta_{mn}$, summing over $P^2$ terms with both $P$'s on the same link (since $\delta_{mn}$ is diagonal, and with the same weight for links in different directions, since $\delta_{00} = \delta_{11} = +1$). However, $\epsilon_{mn}$ is off-diagonal, and so multiplies two $P$'s on different links with a common vertex. (This is obvious in the case of a square lattice.) Its antisymmetry produces a commutator of the fields in $S_2$ [10]. Note that the $\epsilon_{mn}$ term has an additional factor of "$-i$" on the *Euclidean* world sheet as compared to the Minkowski world sheet. This factor goes naturally with the commutator of the fields, which is antihermitian (if the fields are hermitian; if the fields are elements of a Lie algebra, the commutator gives an $if_{ijk}$). There is also an $i$ multiplying the $(\partial X)P$ term in $S_1$ so that it gives the usual Fourier transform when used as $e^{S_1'}$.

A further restriction that must be applied to this $S_1$, but which is not obvious from considerations of $S_P$, is the world-sheet chirality (holomorphy) of the currents $P$. For this purpose we construct $S_1$ by starting in the Hamiltonian formalism with an affine Lie algebra that is a suitable generalization of the algebra of covariant derivatives for the particle, as reviewed in the previous section. By comparison with the form of $S_1$ given above, this requires that $B_{AB} \equiv E_A{}^M E_B{}^N B_{MN}$ be constant. The Virasoro operators are then determined by the usual (classical) current algebra construction. After including any additional first- and second-class constraints, the Hamiltonian form of the action is then completely determined, and leads directly to the first- and second-order forms.

## 4. SUPERSTRINGS

This type of contruction was already applied to the superstring in [3,4], but the representation of the algebra used was not suitable for deriving an action of the form $S_1$ above because there was not an invertible metric on the coordinates, although the metric on the algebra itself was invertible. Green [7] solved this problem by using a representation (the one following from the standard group theory construction) with a coordinate for each current, so the metric on the coordinates became the same as the one on the algebra.

The ordinary Lie algebra $T_A = (d_\alpha, p_a, \omega^\alpha)$ with nonvanishing commutators

$$\{d_\alpha, d_\beta\} = -2\gamma^a_{\alpha\beta} p_a, \quad [d_\alpha, p_a] = -2\gamma_{a\alpha\beta}\omega^\beta$$

has two representations in terms of group-space coordinates $\theta^\alpha, x^a, \phi_\alpha$, corresponding to left and right group multiplication. We can also define a group metric (not the Cartan metric, which vanishes):

$$tr(T_A T_B) = \eta_{AB} = \begin{pmatrix} 0 & 0 & \delta^\beta_\alpha \\ 0 & \eta_{ab} & 0 \\ -\delta^\alpha_\beta & 0 & 0 \end{pmatrix}, \quad \eta^{AB} = \begin{pmatrix} 0 & 0 & -\delta^\alpha_\beta \\ 0 & \eta^{ab} & 0 \\ \delta^\beta_\alpha & 0 & 0 \end{pmatrix}$$

From the representation of the group elements

$$U = e^{-i\phi_\alpha T^\alpha} e^{-ix^a T_a} e^{-i\theta^\alpha T_\alpha}$$

(chosen asymmetric for simplicity with the chosen chirality of the constraints), we find the explicit expressions

$$J^\alpha = \partial\theta^\alpha, \quad J^a = \partial x^a - i(\partial\theta^\alpha)\gamma^a_{\alpha\beta}\theta^\beta, \quad J_\alpha = \partial\phi_\alpha - 2i(\partial x^a)\gamma_{a\alpha\beta}\theta^\beta - \tfrac{2}{3}(\partial\theta^\beta)\gamma^a_{\beta\gamma}\theta^\gamma\gamma_{a\alpha\delta}\theta^\delta$$

The finite form of these differentials is

$$\theta_1 - \theta_2, \quad (x_1 - x_2) - i\theta_1\gamma\theta_2, \quad (\phi_1 - \phi_2) - i(\not{x}_1 - \not{x}_2)(\theta_1 + \theta_2) - \tfrac{1}{3}(\theta_1\gamma\theta_2)\cdot\gamma(\theta_1 + \theta_2)$$

Similar expressions can be obtained for the right-handed currents:

$$\tilde{J}^\alpha = \partial\theta^\alpha, \quad \tilde{J}^a = \partial x^a + i(\partial\theta^\alpha)\gamma^a_{\alpha\beta}\theta^\beta, \quad \tilde{J}_\alpha = \partial\phi_\alpha - 2i(\partial\theta^\beta)\gamma_{a\alpha\beta}x^a - \tfrac{2}{3}(\partial\theta^\beta)\gamma^a_{\beta\gamma}\theta^\gamma\gamma_{a\alpha\delta}\theta^\delta$$

From these we can read off the expressions for $E$ and $\tilde{E}$ and invert them, so we can also write the covariant derivatives:

$$d_\alpha = i\partial_\alpha - \gamma^a_{\alpha\beta}\theta^\beta\partial_a - i\tfrac{4}{3}\gamma^a_{\alpha\beta}\theta^\beta\gamma_{a\gamma\delta}\theta^\gamma\partial^\delta, \quad p_a = i\partial_a - 2\gamma_{a\alpha\beta}\theta^\alpha\partial^\beta, \quad \omega^\alpha = i\partial^\alpha$$

$$\tilde{d}_\alpha = i\partial_\alpha + \gamma^a_{\alpha\beta}\theta^\beta\partial_a - 2x^a\gamma_{a\alpha\beta}\partial^\beta + i\tfrac{2}{3}\gamma^a_{\alpha\beta}\theta^\beta\gamma_{a\gamma\delta}\theta^\gamma\partial^\delta, \quad \tilde{p}_a = i\partial_a, \quad \tilde{\omega}^\alpha = i\partial^\alpha$$

Since $\{\omega,\omega\} = \{\tilde{\omega},\tilde{\omega}\} = 0$ (in fact, $\omega = \tilde{\omega}$), we can apply $\tilde{\omega} \equiv i\partial/\partial\phi = 0$ as a first-class constraint to eliminate $\phi$ and get the usual superspace. Then $p = \tilde{p}$ is the usual momentum. We can also identify $d$ as the usual supersymmetry covariant spinor derivative, while $q \equiv \tilde{d}$ is the supersymmetry generator. Thus, as in the usual group theory construction that starts with just $d$ and $p$, $d$ and $p$ are identified with supersymmetry covariant derivatives, while $\tilde{d}$ and $\tilde{p}$ are identified with the supersymmetry generators themselves.

Even before eliminating $\phi$, we find spacetime translations $\tilde{p}$ represented by simple translations of $x^a$. Similarly, we have $\tilde{\omega}$ generating simple translations of $\phi$. However, supersymmetry transformations act nontrivially on $\phi$:

$$\delta\theta^\alpha = \epsilon^\alpha, \quad \delta x^a = -i\epsilon^\alpha\gamma^a_{\alpha\beta}\theta^\beta, \quad \delta\phi_\alpha = 2i\epsilon^\beta\gamma_{a\alpha\beta}x^a + \tfrac{2}{3}(\epsilon^\beta\gamma^a_{\beta\gamma}\theta^\gamma)\gamma_{a\alpha\delta}\theta^\delta$$

For the superstring we can apply the general expressions for the covariant derivatives, reviewed in section 2. All the terms can be read directly from the above expressions for the particle, except for the $B$ terms. For this case, in an appropriate gauge for $B$, $B_{AB}$ can be written surprisingly as a constant matrix (though $B_{MN}$ and $\tilde{B}_{AB}$ cannot):

$$B_{MN} = \begin{pmatrix} 0 & i\gamma_{n\mu\sigma}\theta^\sigma & \tfrac{1}{2}\delta^\beta_\alpha \\ -i\gamma_{m\nu\sigma}\theta^\sigma & 0 & 0 \\ \tfrac{1}{2}\delta^\alpha_\beta & 0 & 0 \end{pmatrix} \quad \Rightarrow \quad B_{AB} = \tfrac{1}{2}\begin{pmatrix} 0 & 0 & \delta^\beta_\alpha \\ 0 & 0 & 0 \\ \delta^\alpha_\beta & 0 & 0 \end{pmatrix}$$

($B_{AB}$ is graded antisymmetric, so this part is symmetric.) Our choice of gauge for $B$ under the usual gauge transformation $\delta B_{MN} = \partial_{[M}\lambda_{N)}$, generated by the unitary transformations $\ln U = i\int d\sigma\, X'^M \lambda_M$, differs from the usual "chiral" representation by $\lambda^\alpha \sim \theta^\alpha$ or $\lambda_\alpha \sim \phi_\alpha$ ($\int X'^M\lambda_M \sim \int \theta'^\alpha\phi_\alpha$). The only change in $B_{MN}$ is the $\delta$ terms.

The resulting representations for the affine Lie algebra with constant $B_{AB}$ are:

$$D_\alpha = (i\delta_\alpha - \tfrac{1}{2}\phi'^\alpha) + i\gamma^a_{\alpha\beta}\theta^\beta(i\delta_a + x'_a) + \tfrac{1}{3}[(4i\delta^\gamma + \theta'^\gamma)\gamma^a_{\gamma\delta}\theta^\delta]\gamma_{a\alpha\beta}\theta^\beta$$

$$P_a = (i\delta_a + x'_a) - i(2i\delta^\alpha + \theta'^\alpha)\gamma_{a\alpha\beta}\theta^\beta$$

$$\Omega^\alpha = i\delta^\alpha + \tfrac{3}{2}\theta'^\alpha$$

$$\tilde{D}_\alpha = (i\delta_\alpha + \tfrac{3}{2}\phi'^\alpha) - i\gamma^a_{\alpha\beta}[\theta^\beta i\delta_a - 2i\delta^\beta x_a + (\theta^\beta x_a)'] - \tfrac{2}{3}(i\delta^\gamma\gamma^a_{\gamma\delta}\theta^\delta)\gamma_{a\alpha\beta}\theta^\beta$$

$$\tilde{P}_a = i\delta_a - x'_a$$

$$\tilde{\Omega}^\alpha = i\delta^\alpha - \tfrac{1}{2}\theta'^\alpha$$

On the other hand, in the chiral representation $\tilde{\Omega} = i\delta/\delta\phi$, and there is no $\phi'$ term in $D$. In that representation the original representation of the $D, P, \Omega$ algebra [3] is obtained simply by dropping all $\delta/\delta\phi$ terms.

As explained above, the zero modes of $D$ and $\tilde{D}$ cannot both be considered supersymmetries. However, if we again impose the first-class constraint $\tilde{\Omega} = 0$, since $\{\tilde{D}, \tilde{\Omega}\} \sim 2i\delta'$, all but the zero mode of $\tilde{D}$ can be gauged away. (The non-zero modes of $\tilde{D}$ are canonical conjugates of the non-zero modes of $\tilde{\Omega}$, and the conjugates of first-class constraints are gauge degrees of freedom.) This leaves $q \equiv \int d\sigma \tilde{D}$ as the only surviving part of $\tilde{D}$, so it can again be interpreted as supersymmetry. (Also, all of $\tilde{P}$ survives, and is used to define the right-handed Virasoro operators.) However, before applying the constraint $\tilde{\Omega} = 0$, the zero modes of $\tilde{D}, \tilde{P}, \tilde{\Omega}$ have no $X'^M$ terms, and have the same form as in the particle case. This is because the action expressed in terms of $D, P, \Omega$ with constant $B_{AB}$ is invariant under these symmetries without total derivative terms. (In the old representation, the total derivative terms are responsible for generating the $X'$ terms in $q$ by the Noether procedure [10].) In addition to $\tilde{\Omega}$, we also have as first-class constraints the left- and right-handed Virasoro operators $\frac{1}{2}\eta^{BA} P_A P_B$ and $\frac{1}{2}\eta^{BA} \tilde{P}_A \tilde{P}_B$, and the second-class constraint $D$. This describes the heterotic string in a formulation equivalent to the Green-Schwarz one. The generalization to the other superstrings is straightforward, by introducing two $\theta$'s and imposing opposite-(world-sheet-)chirality constraints on them.

## 5. NEW ACTION

As a result of this construction the Green-Schwarz action can be rewritten in the very simple second-order form

$$L_2 = -\frac{1}{2}\mathbf{g}^{mn} J_n^a J_{ma} - i\frac{1}{2}\epsilon^{mn} J_n^\alpha J_{m\alpha}$$

This is easy to check by direct evaluation: The $\mathbf{g}$ term is the usual $\mathbf{g}$ term of the Green-Schwarz action. For the $\epsilon$ term, the $(\theta\gamma\partial\theta)^2$ term vanishes by symmetry, while the $(\partial\theta)(\partial\phi)$ term vanishes by integration by parts. The remaining term in the Wess-Zumino term is identical to the Green-Schwarz Wess-Zumino term. This expression is simpler than the Green-Schwarz one in the sense that it is quadratic in the manifestly supersymmetric $J$'s, while for the Green-Schwarz action the Wess-Zumino term must be written with an unphysical third "world-sheet" dimension to be manifestly covariant (just as here we used an unphysical coordinate $\phi$), and is then cubic. This is a consequence of our special choice of gauge where $B_{AB}$ is constant:

The total derivative term $\epsilon(\partial\theta)(\partial\phi)$ in $L_2$ results from the gauge transformation from the standard gauge to our gauge.

This translates into first-order form as

$$L_1 = i(\partial_m X^M)E_M{}^A P_A^m - \tfrac{1}{2}\mathbf{g}_{mn}P^{na}P_a^m + 2i\epsilon_{mn}P^{n\alpha}P_\alpha^m$$

Except for the appearance of $P^{1a}$, this is also the Hamiltonian form: $P^{1a}$ appears quadratically, and can be integrated out trivially. On the other hand, $P^{1\alpha}$ and $P_\alpha^1$ appear linearly, as Lagrange multipliers enforcing the constraints $D_\alpha = 0$ and $\tilde{\Omega}^\alpha = 0$, which were described in the previous section. (These constraints also kill the terms $\Omega^\alpha D_\alpha$ and $\tilde{\Omega}^\alpha \tilde{D}_\alpha$ in the Virasoro operators.)

The reason that an action formally identical to the Green-Schwarz action (at least in second-order form) can be written in a form where the Lagrangian is manifestly supersymmetric without total derivative terms is that we were able to drop $\phi$ dependence only after integration by parts. Although the fact that the gauge transformation that removes $\phi$ is Abelian and nonderivative means $\phi$ plays a trivial role in the classical mechanics, it clearly plays a nontrivial one in the lattice quantization, and may also do so in the continuum quantization: Its gauge fixing might be related to the nontrivial zero modes associated with $\kappa$ symmetry, which are needed to produce vertex factors in Green-Schwarz amplitudes, even in light-cone gauges [13]. Also, the appearance of $\phi$ is the result of a $B$ gauge transformation, and the existence of a gauge where $B_{AB}$ is a constant must have some special geometrical or group theoretical significance. At least, $B_{AB}$ should be considered as a group metric, like $G_{AB}$, which is also a constant only in a certain gauge for tangent-space gauge transformations [14].

Since the interaction term appears exponentially in the functional integral (as do all terms in the action $S_1'$), the vertex operator appears exponentially. We can then write the particle field theory action as

$$S_P = tr \int d\theta\, dx\, d\phi \left( -\tfrac{1}{2}\psi e^{p_a^2/2}\psi + \tfrac{1}{6}\frac{G}{\sqrt{N}}\psi e^{2i\omega_1^\alpha d_{2\alpha}}\psi\psi \right)$$

where we have chosen to use three-point interactions. (The explicit expressions for $d, p, \omega$ were given in the previous section.) The labels "1" and "2" on $\omega$ and $d$ are to indicate that for each term in the expansion of the exponential we are to sum (with appropriate signs) over the two combinations of $\omega$ and $d$ acting on different $\psi$'s.

Assuming it is sufficient to keep just the leading terms when using the 1/N expansion, the action becomes

$$S_P \approx tr \int d\theta\, dx\, d\phi \left[ -\tfrac{1}{2}\psi(1 + \tfrac{1}{2}p_a^2)\psi + \frac{G}{\sqrt{N}}(\tfrac{1}{6}\psi^3 + i\tfrac{1}{3}\psi\{\omega^\alpha\psi, d_\alpha\psi\}) \right]$$

This action is partially gauge fixed, and will need more investigation to find both the gauge invariant action and the completely gauge fixed one. Since only world-sheet reparametrizations have been fixed by the random lattice, both the usual gauge invariances associated with supersymmetry (such as $\kappa$ invariance) and the new gauge invariance associated with $\phi$ need gauge fixing in both $S_1$ and $S_P$. However, it is important to first understand the gauge invariant action, and there may be a direct relation between those invariances in $S_1$ and $S_P$. Since $\kappa$ symmetry relates the $g$ and $\epsilon$ terms in $S_1$, it relates the kinetic and interaction terms in $S_P$. This was to be expected, since this symmetry is generated by the field equation $\slashed{p}d = 0$, which gets extra contributions from the interactions. Another possibility is that the particle field theory, since it is massive, might not need gauge invariances, if it lacks Stueckelberg fields. Also, if we consider a compactification to four dimensions with N=1 supersymmetry, instead of the ten-dimensional superstring, the particle quantization might explain the string quantization, since 4D N=1 supergraphs are already well understood.

We have written the particle field theory action for the heterotic string, but it generalizes straightforwardly to other superstrings. However, the heterotic case may be more interesting, since in D=10 the only known superparticle theory whose fields can carry a group theory index (to be NxN matrices) is super Yang-Mills theory, which exists there only as an N=1 supersymmetric theory. In that case we also need chiral bosons on the world sheet, to compensate for the difference in left- and right-handed critical dimensions. A form of their Lagrangian [15] that can be latticized by these methods is

$$-\tfrac{1}{2}\mathbf{g}^{mn}(\partial_m\chi)(\partial_n\chi) - \tfrac{1}{2}\lambda_{mn}(\mathbf{g}^{mp} - i\epsilon^{mp})(\mathbf{g}^{nq} - i\epsilon^{nq})(\partial_p\chi)(\partial_q\chi)$$

## ACKNOWLEDGMENTS

I thank Nathan Berkovits and Martin Roček for discussions. This work was supported in part by the National Science Foundation Grant No. PHY 9309888.